\documentstyle[12pt]{article}

\def\zm{M_{_Z}}
\def\gz{\Gamma_{_Z}}
\def\zms{M_{_Z}^2}

\def\hm{m_{_H}}

\def\s0h{\sigma^0_h}

\def\gh{\Gamma_h}
\def\gb{\Gamma_b}

\def\g2{g^2}

\def\ord{{\cal O}}

\def\osp2{16\,\pi^2}

\def\ap2{\left(p^2\right)}

\def\ap{{{\alpha} \over {\pi}}}

\begin{document}

\rightline{CERN-TH.7463/94}
\vspace{0.5cm}

\noindent
{\Large{\bf {\tt TOPAZ0}~2.0 - A program for
computing de-convoluted and realistic observables
%and for fitting cross sections and forward-backward asymmetries
around the {\boldmath $Z^0$} peak }}

\vspace{0.5cm}
\medskip
\noindent
Guido MONTAGNA$^a$, Oreste NICROSINI$^{b,}$\footnote{\footnotesize {\it On
leave
from INFN, Sezione di Pavia, Italy}}, Giampiero PASSARINO$^c$ and
Fulvio PICCININI$^d$ \\

\medskip
\noindent
{\it $^a$Dipartimento di Fisica Nucleare e Teorica, Universit\`a di
         Pavia, via A. Bassi n. 6 - 27100 PAVIA - ITALY} \\
\noindent
{\it $^b$Theory Division, CERN CH-1211, GENEVA 23 - SWITZERLAND} \\
\noindent
{\it $^c$Dipartimento di Fisica Teorica, Universit\`a di Torino, and INFN,
Sezione di Torino, via P. Giuria n. 1 - 10125 TURIN - ITALY} \\
\noindent
{\it $^d$ INFN, Sezione di Pavia, via A. Bassi n. 6 -
27100 PAVIA - ITALY}

\bigskip
\noindent
Program classification: 11.1 \\
\bigskip

\noindent
{\small
The program {\tt TOPAZ0} is designed for computing
%and fitting
$Z^0$ parameters,
de-convoluted and QED-dressed cross sections
and forward-backward asymmetries of $e^+ e^-$ annihilation
into fermion pairs
and of Bhabha scattering around the $Z^0$ peak, over both a completely
inclusive experimental set-up
and a realistic one, i.e. with cuts on acollinearity, energy or invariant
mass and angular acceptance of the outgoing fermions. The new version,
 2.0,
offers the possibility of imposing different experimental cuts on cross
sections and forward-backward asymmetries in a single run,
%improves the treatment of
%the experimental error according to the most recent literature
 and includes radiative
corrections whose effect can become relevant in view of the present and
foreseen experimental accuracy.
Moreover, an additional option is included,
which allows an estimate of the theoretical uncertainty due to
unknown higher-order effects, both of electroweak and QCD origin.
 With respect to the version 1.0,
the code is available in the
form of {\tt SUBROUTINE}, in order to render more viable
the use of the program for aims not planned by the {\tt TOPAZ0} package
itself. }

\begin{center}
Submitted to Computer Physics Communications \\
\end{center}

\vfil

\leftline{CERN-TH.7463/94}
\leftline{June 1995}
\eject

\leftline{\Large{\bf NEW VERSION SUMMARY}}
\vskip 15pt

\leftline{{\it Title of new version:} {\tt TOPAZ0}~2.0}
\vskip 8pt

\leftline{{\it Catalogue number:} }
\vskip 8pt

\noindent
{\it Program obtainable from:} CPC Program Library, Queen's
University of Belfast, N. Ireland (see application form in this issue)
\vskip 8pt

\noindent
{\it Reference to original program:} {\tt TOPAZ0}; {\it Cat.~no.:} ACNT;
{\it Ref. in CPC: } 76 (1993) 328
\vskip 8pt

\noindent
{\it Authors of original program:} Guido Montagna, Oreste Nicrosini,
Giampiero Passarino, Fulvio Piccinini and Roberto Pittau
\vskip 8pt

\noindent
{\it The new version supersedes the original program}
\vskip 8pt

\noindent
\leftline{{\it Licensing provisions:} none}
\vskip 8pt

\noindent
{\it Computer for which the new version is designed:}
VAX, HP-APOLLO 7000; {\it Installation:} INFN,
Sezione di Pavia, via A.~Bassi 6, 27100 Pavia, and Sezione di Torino,
via P.~Giuria 1, 10125 Turin, Italy
\vskip 8pt

\leftline{{\it Operating system under which the new version has been tested:}
VMS, UNIX}
\vskip 8pt

\leftline{{\it Programming language used in the new version:} FORTRAN 77}
\vskip 8pt

\noindent
\leftline{{\it Memory required to execute with typical data} }
\noindent
300-600 kbyte as evaluator of observables in seven energy points
\vskip 8pt

\leftline{{\it No. of bits in a word: } 32}
\vskip 8pt

\leftline{{\it No. of processors used: } 1}
\vskip 8pt

\leftline{\it The code has not been vectorized}
\vskip 8pt

\leftline{{\it Subprograms used:} NAGLIB~[1] }
\vskip 8pt

\leftline{{\it No. of lines in distributed program, including test data,
etc.:} 12343 }
\vskip 8pt

\leftline{\it Correspondence to:}
\leftline{GIAMPIERO@TO.INFN.IT, NICROSINI@PV.INFN.IT }
\leftline{VAXTO::GIAMPIERO, VAXPV::NICROSINI}
\vskip 8pt

\noindent
{\it Keywords:} $e^+ e^-$ annihilation, Bhabha scattering, LEP,
$Z^0$ resonance, electroweak, extrapolated and realistic experimental set-up,
QCD corrections, QED corrections, pure weak corrections, radiative corrections,
Minimal Standard Model, de-convoluted and
realistic observables,
%fit to cross sections and forward-backward asymmetries,
\\ theoretical uncertainties.
\vskip 8pt

\leftline{{\it Nature of physical problem} }
\noindent
An accurate theoretical description of $e^+ e^-$ annihilation
processes and of Bhabha scattering at the $Z^0$ resonance is necessary in
order to compare theoretical cross sections and asymmetries with the
experimental ones as measured by the LEP collaborations. In particular a
{\it realistic } theoretical description, i.e. a description in which the
effects of experimental cuts, such as maximum acollinearity, energy or
invariant mass and angular acceptance of the outgoing fermions, are taken into
account, allows
the comparison of the Minimal Standard Model predictions with
%the unknown parameters of the , $M_Z$, $m_t$, $m_H$ and
%$\alpha_s$, to be fitted over
experimental {\it raw} data, i.e. data corrected for
detector efficiency but not for acceptance. The program takes
into account
all the corrections, pure weak, QED and QCD, which allow for such a
{\it realistic} theoretical description.
\vskip 8pt

\leftline{{\it Method of solution} }
\noindent
Same as in the original program.
A detailed description of the theoretical formulation and of a sample of
physical results obtained can be found in~[2].

\vskip 8pt

\leftline{\it Reasons for the new version}
\noindent
The new version gives the possibility of computing
%and fitting
observables in
an experimental set-up with different cuts on cross sections and asymmetries,
according to the most recently published LEP data.
%The experimental error
%treatment in the fitting branch has been improved according to the most
%recent literature.
Radiative corrections whose effect
can become relevant in view of the present and
foreseen experimental accuracy have been included. An option is added which
allows an estimate to be made of
the theoretical uncertainty associated with unknown higher-order
radiative corrections of electroweak and QCD origin.
\vskip 8pt

\leftline{{\it Restrictions on the complexity of the problem} }
\noindent
The theoretical formulation is specifically worked out for energies around
the $Z^0$ peak.
Analytic formulas have been developed for an experimental set-up with
symmetrical angular acceptance. Moreover the angular acceptance of the
scattered antifermion has been assumed to be
larger than the one of the scattered
fermion. The prediction for Bhabha scattering is understood to be for the
large-angle regime.

\vskip 8pt

\leftline{{\it Typical running time} }
\noindent
Dependent on the required experimental set-up. As evaluator of observables
in seven energy points, between 10 (extrapolated set-up) and 270 (realistic
set-up) CPU seconds for HP-APOLLO 7000, corresponding to 100-2500 CPU
seconds for VAX 6410.
%As fitter, the running time is roughly given by the
%time needed as evaluator times the number of iterations requested to obtain
%convergence: dependent on the data sample.
As estimator of the theoretical
uncertainty, it requires about 5400 CPU seconds for HP-APOLLO 7000,
for a single energy point.
\vskip 8pt

\leftline{{\it Unusual features of the program} }
\noindent
Subroutines from the library of mathematical subprograms NAGLIB
%both
for the numerical integrations
%and the minimization procedure
are used in the program.
\vskip 8pt

\leftline{{\it References} }
\noindent
[1] {NAG Fortran Library Manual Mark 15 (Numerical Algorithms
Group, Oxford, 1991).}
\vskip 10pt\noindent
[2] {G.~Montagna, O.~Nicrosini, G.~Passarino, F.~Piccinini and
R.~Pittau, Nucl.~Phys. B401 (1993) 3. }
\vskip 15pt
\vfil \eject

\leftline{\Large{\bf LONG WRITE-UP}}
\vskip 15pt

\noindent
\section{Introduction}
\vskip 10pt

The high precision reached by the four LEP experiments has motivated several
groups in assembling Monte Carlo programs or semi-analytical codes for
giving accurate theoretical predictions relevant in the region around
the $Z^0$ peak~[1-6].
As a consequence we have by now several electroweak libraries
for radiative corrections at LEP energies. This is a very important fact
because, when precision physics is the main goal, continuous
cross-checks are needed, especially at the moment when several new effects
have been computed, due for instance to QCD corrections or large-$m_t$
behaviour~\cite{yrwg}.

%At any rate, building a fully self-consistent electroweak library is only one
%of the reasons that stimulated us to write {\tt TOPAZ0}.
%It is actually our opinion that the simple evaluation of physical quantities,
%both on inclusive and realistic experimental set-up, cannot be
%the final task and an attempt should be made to analyse directly
%the experimental data.

%-- {\bf modifica} --

As it stands {\tt TOPAZ0} can be used to compute
$Z^0$ parameters and de-convoluted observables but also
%to perform
%fits in a self-consistent way, since no additional software is needed in this
%case.
 to obtain predictions for QED-dressed distributions over a
realistic set-up resembling the experimental {\it raw} data.
 The new version, {\tt TOPAZ0}~2.0,
offers improvements both in the direction of
computing observables in a realistic set-up
according to the selection criteria
of the most recently published experimental data
%and fitting the most recently
%published experimental data
and in the direction of taking into account
radiative corrections whose effect can become relevant in view of the
experimental accuracy reached at present and foreseen for the near future.
As far as the first kind of improvements is concerned,
{\tt TOPAZ0}~2.0 allows one to put
different cuts on the cross section and the forward-backward asymmetry in a
given channel, thus matching the set-up adopted by some of the LEP
collaborations.
%moreover the fitting branch has been upgraded to take into
%account a proper treatment of correlated experimental errors according to
%the most recent literature.
As far as the second kind of improvements is
concerned, all recently computed electroweak and QCD corrections are
implemented. Moreover, further QED corrections whose size can become
comparable with the foreseen experimental accuracy are included.

For the time being, a particularly important item is to estimate the
uncertainty intrinsic to the theoretical predictions. For instance as a
consequence of the high experimental accuracy reached by the LEP
collaborations, the theoretical uncertainty affects the determination
of the unknown parameters of the standard model~\cite{lettera}.
To this aim, a new feature has been implemented in {\tt TOPAZ0},
namely the possibility of running the
evaluator over several theoretical options, related to different
implementations
of higher orders in the perturbative expansion.
As a result, when choosing
the proper branch in {\tt TOPAZ0}, the code returns an estimate of the
theoretical uncertainty of electroweak and QCD origin.

%-- {\bf modifica} --

With respect to the previous version, the code has been rearranged
in the form of {\tt SUBROUTINE}, in order to render more viable
the use of the program for aims not planned by the {\tt TOPAZ0} package
itself.

\noindent
\section{The most important new features}
\vskip 10pt

In the following we list and shortly comment the most important modifications
of {\tt TOPAZ0}. They can be classified into three classes.

\vskip 24pt \noindent
TECHNICAL MODIFICATIONS

\begin{itemize}

\item The present version allows the experimental cuts
on cross sections and asymmetries in a given leptonic channel
to be treated separately, according to
the most recently adopted experimental strategy. To this aim the routines
{\tt OBSERVABLES } and {\tt OBSCUT} have been updated. For instance the
present version allows the computation of an extrapolated cross section
and a cut asymmetry in the same run.

%\item In the fitting branch the error treatment has been improved following
%      the error matrix approach~\cite{errori}.

\item The new flag {\tt OTHERR} has been included. For {\tt OTHERR = Y} the
      evaluator runs over several ($2^7$) theoretical options, which reflect
      different (but not antithetic) implementations of radiative corrections,
      leading to  a different treatment of missing higher-order terms.
      For {\tt OTHERR = Y} the program returns a separate estimate of the
      weak and QCD uncertainty on the one hand and of the electromagnetic
      one on the other hand.

%-- {\bf modifica} --

%{\sl

\item The code has been prepared in the following form of {\tt SUBROUTINE}:
\begin{verbatim}
SUBROUTINE TOPAZ0(NRST,TRS,TZM,TTQM,THM,TALS)
\end{verbatim}
where \\
{\tt NRST}: the number of centre of mass energies \\
{\tt TRS}: the values of centre of mass energy (GeV) \\
{\tt TZM}: $Z^0$ mass (GeV) \\
{\tt TTQM}: top-quark mass (GeV) \\
{\tt THM}: Higgs boson mass (GeV) \\
{\tt TALS}: $\alpha_s(M_Z)$ \\
are the basic input parameters. Moreover, the {\tt SUBROUTINE TINPUT}
is provided in order to
allow the user to supply the experimental cuts for a given
$Z^0$ decay channel and to select among different theoretical options.
The meaning of the latter input parameters/flags is summarized
at the beginning of {\tt SUBROUTINE TINPUT}.
%--{\bf modifica}--
In particular, as input/output facility, the flag {\tt OMON} allows to
write down the results
for the de-convoluted and realistic observables ({\tt OMON = Y}),
or to store them into appropriate common blocks {\tt /TTH, /TPO, /TPWTH}
({\tt OMON = N}).
This second choice is supplied for fitting purposes.

\end{itemize}

\vskip 24pt \noindent
THEORETICAL IMPROVEMENTS: \\
PURE WEAK AND QCD CORRECTIONS

An important fact to realize is that there were several new calculations of
radiative corrections in the last couple of years: their effect must be
understood, classified descriptively, systematized and codified. Here it will
only be possible to present a brief summary list of some of these effects:

\begin{itemize}

\item By using a dispersion relation the value of $\alpha(\zm)$ obtained
      from experimental data is now $\alpha(\zm) =
      (128.87 \pm 0.12)^{-1}$~\cite{je}.
      %-- {\bf modifica} -- {\sl Citerei Jegerlehener come da
      %CERN YR EWWG Report (TASI 1990), i.e.~\cite{je}. Non \`e il caso
      %di mettere una footnote sulle tre recenti analisi, come nel YR? }

\item Two-loop heavy top effects in $\rho, g_{V,A}(b)$ for arbitrary values
      of $\hm$~\cite{barb}.

\item $\ord(\alpha\alpha_s)$ final state radiation~\cite{kat}.

\item $\ord(\alpha\alpha_s)$ corrections to vector boson
self-energies~\cite{aas}.

\item Complete $\ord(\alpha_s^2{{m_b^2}\over {\zms}})$ corrections to
      $\Gamma_A$, including NNLO terms for the running b-quark
mass~\cite{qcd}.

\item Complete $\ord(\alpha_s^2)$ corrections to $\gz$ and $\ord(\alpha_s^3)$
      corrections to $\gh$~\cite{qcd}.

\item $\ord(\alpha_s G_F m_t^2)$ corrections to $\gb$, the FJRT
effect~\cite{ftjr}.

\item The $\ord(\alpha\alpha_s^2)$ correction to $\rho$~\cite{tar}.

\end{itemize}

To illustrate the numerical relevance of those effects included in
our electroweak library we can compare the peak cross sections and the
forward-backward asymmetries for $\mu, \tau$ and hadronic channels in a
fully extrapolated set-up and for the same input parameters as used
in~\cite{topaz0cpc}~(see table~1).

\begin{table}[htbp]
\begin{center}
\begin{tabular}{|c|c|c|}
\hline
% $\sqrt{s} (GeV)$ & \multicolumn{3}{c|}
% {} \\
% \cline{2-4}
               Results   & TOPAZ0   & TOPAZ0 2.0         \\
\hline
                             &          &                    \\
             $\sigma_{\mu}$   &  1.4912  & 1.4900     \\
                             &          &                    \\
             $\sigma_{\tau}$  &  1.4836  & 1.4824     \\
                             &          &                    \\
             $\sigma_{had}$   &  30.615  & 30.591     \\
                             &          &                    \\
             $A_{FB}(\mu)$    &  0.0047  & 0.0039     \\
                             &          &                    \\
             $A_{FB}(\tau)$   &  0.0047  & 0.0039     \\
                             &          &                    \\
\hline
\end{tabular}

\caption{The effect of the new weak and QCD corrections implemented in TOPAZ0
2.0 for  $ E_{c.m.} = 91.222 $~GeV,  $ M_Z = 91.175 $~GeV, $ m_t = 200 $~GeV,
$ m_{H} = 250 $~GeV and $ \alpha_s = 0.125$ }

%{\it
%The effect of the new weak and QCD corrections implemented in TOPAZ0
%2.0 for $\sqrt{s} = 91.222$~GeV,
%$M_Z = 91.175\,$~GeV, $m_t = 200\,$~GeV, $m_{H} = 250\,$~GeV
%and $\alpha_s = 0.125$.
%}
%\label{ta1}

\end{center}
\end{table}
\normalsize

\vfil \eject
\noindent
THEORETICAL IMPROVEMENTS: QED CORRECTIONS

\begin{itemize}

\item The exact ${\cal O}(\alpha)$ initial-final state interference
(including hard brems\-strah\-lung)
has been added
for the leptonic channels as follows. The soft and
virtual contribution has been included as a 1-dimensional spectrum ($d \sigma
/ d \cos \vartheta$, $\vartheta$ being the fermion scattering angle),
taken from~\cite{hollik} and numerically integrated over
the forward and backward hemispheres to build the correction to the cross
section and the asymmetry. For the hard bremmstrahlung contribution, the
5-dimensional spectrum quoted in~\cite{bkj} has been worked out analytically
to obtain a twofold distribution ($d \sigma / d \cos \vartheta d E_\gamma$,
$E_\gamma $ being the photon energy) and then numerically integrated as
above. The analytical work performed on the hard photon part allows the
computation of
 the interference correction with no substantial change in CPU time.
The inclusion of the interference contribution is controlled, for each energy
point, by a new flag {\tt ONIF}. For {\tt ONIF = Y}, the new subroutine {\tt
IFINT } is called and the correction to the cross section and the asymmetry
is returned after calling {\tt FUNSUBS} and {\tt FUNSUBH}
for the soft and hard contribution, respectively.
 It should be pointed out that
 the treatment of the intial-final state interference is exact for
$s-$channel annihilation but approximate for full Bhabha
 scattering.

\item The ${\cal O}(\alpha^2)$ leptonic and hadronic contribution from
initial-state radiation according to~\cite{kkks} has been included as follows.
The
soft and virtual contributions are computed using the analytical formula
given in~\cite{kkks}. For the hard contribution, the 1-dimensional spectrum
of~\cite{kkks} has been integrated numerically. The total relative
contribution to the cross sections in the hadronic and leptonic channels is
returned for each energy point (for {\tt ONP = Y})
after a call to subroutine {\tt PAIRS}, which
in turn calls the subroutines {\tt FUNSUBE}, {\tt FUNSUBMU} and {\tt
FUNSUBHP},
for $e$, $\mu$ and $h$ ${\cal O}(\alpha^2)$ hard contributions respectively.

\end{itemize}

\vskip 15pt

\noindent
\section{Test Run Output}
\vskip 10pt

The typical calculations that can be performed with the new version of the
program are illustrated in the following example.

\begin{verbatim}


 RESIDUAL WEAK CORRECTIONS ARE COMPUTED AT THE PEAK
 PAIR PRODUCTION EFFECT IS INCLUDED FOR EACH ENERGY
 I-F STATE INTERFERENCE IS INCLUDED FOR EACH ENERGY  FOR LEPTONS ONLY
 S+T CHANNEL FOR ELECTRONS ARE COMPUTED
 NO CUT ON THE REDUCED ENERGY IS ASSUMED
 FULL PHASE SPACE FOR FINAL STATE EM RADIATION  IS ASSUMED

  MINIMUM ANGLE OF MU(-) (DEG):                     0.00000E+00
  MINIMUM ANGLE OF MU(+) (DEG):                     0.00000E+00
  MAX. ACOLLINEARITY ANGLE FOR MUONS (DEG):         0.18000E+03
  MINIMUM ANGLE OF TAU(-) (DEG):                    0.00000E+00
  MINIMUM ANGLE OF TAU(+) (DEG):                    0.00000E+00
  MAX. ACOLLINEARITY ANGLE FOR TAUS (DEG):          0.18000E+03
  ENERGY THRESHOLD  FOR ELECTRONS (GEV):            0.10000E+01
  MINIMUM ANGLE OF E(-) (DEG):                      0.40000E+02
  MINIMUM ANGLE OF E(+) (DEG):                      0.00000E+00
  MAX. ACOLLINEARITY ANGLE FOR ELECTRONS (DEG):     0.25000E+02


  CURRENT VALUES FOR THE PARAMETERS ARE:
  Z MASS (GEV)     =  0.91189E+02       TOP MASS (GEV) =  0.17400E+03
  HIGGS MASS (GEV) =  0.30000E+03       ALPHA_S        =  0.12400E+00

  W MASS    (GEV)       =         0.802951E+02
  NU                    =         0.167156E+00
  ELECTRON              =         0.839136E-01
  MUON                  =         0.839129E-01
  TAU                   =         0.837222E-01
  UP                    =         0.300439E+00
  DOWN(STRANGE)         =         0.383139E+00
  CHARM                 =         0.300387E+00
  BOTTOM                =         0.376022E+00

  SIN^2(E)              =         0.232208E+00
  SIN^2(B)              =         0.233490E+00
  A_FB(L) EFF.          =         0.150042E-01
  A_LR EFF.             =         0.141608E+00
  TOTAL WIDTH (GEV)     =         0.249613E+01
  G_H/G_E               =         0.207727E+02
  SIGMA0_H  (NB)        =         0.414425E+02
  G(B)/G(HAD)           =         0.215718E+00
  A_FB(B)               =         0.991510E-01
  HADRONIC WIDTH (GEV)  =         0.174312E+01
  A^POL_FB(B)           =         0.934252E+00
  A_FB(C)               =         0.706791E-01
  G(C)/G(HAD)           =         0.172328E+00



  DECONVOLUTED FB-ASYMMETRIES ARE:
 Z-EXCHANGE ONLY

  ELECTRON             =        0.1500028E-01
  MUON                 =        0.1500028E-01
  TAU                  =        0.1497573E-01
  CHARM                =        0.7063442E-01
  BOTTOM               =        0.9875908E-01


  DECONVOLUTED FB-ASYMMETRIES ARE:
 COMPLETE

  ELECTRON             =        0.1628186E-01
  MUON                 =        0.1628186E-01
  TAU                  =        0.1629466E-01
  CHARM                =        0.6814782E-01
  BOTTOM               =        0.9661819E-01



  OBSERVABLES ARE

  E_CM (GEV)            =          0.91300E+02

  SIGMA(E)   (NB)       =        0.1195475E+01 +/-        0.1745525E-03
  SIGMA(MU)  (NB)       =        0.1486515E+01 +/-        0.8608965E-04
  SIGMA(TAU) (NB)       =        0.1479019E+01 +/-        0.8288089E-04
  SIGMA(HAD) (NB)       =        0.3051606E+02
  A_FB(E)               =        0.1421349E+00 +/-        0.1666541E-03
  A_FB(MU)              =        0.7492082E-02 +/-        0.7530680E-05
  A_FB(TAU)             =        0.7519941E-02 +/-        0.7543220E-05
  A_FB(C)               =        0.6317556E-01
  A_FB(B)               =        0.9483699E-01


\end{verbatim}

\vskip 15pt

\begin{thebibliography}{99}

\bibitem{zfitter} {D.~Bardin et al., Program {\tt ZFITTER},
CERN preprint CERN-TH.6443/92;
Nucl.~Phys. B351 (1991) 1; Z.~Phys. C44 (1989) 493;
Phys. Lett. B255 (1991) 290.}

\bibitem{bhm}{G.~Burgers, W.~Hollik and M.~Martinez, Program {\tt BHM}. }

\bibitem{leptop}{V.~A.~Novikov, L.~B.~Okun, A.~N.~Rozanov and M.~I.~Vysotsky,
Program {\tt LEPTOP}, CERN preprint CERN-TH.7217/94.}

\bibitem{ali}{W.~Beenakker, F.~A.~Berends and S.~C.~van der Mark, Nucl.~Phys.
B349 (1991) 323.}

\bibitem{z0polo}{B.~A.~Kniehl and R.~G.~Stuart, Comput. Phys. Commun. 72 (1992)
175.}

\bibitem{topaz0cpc} {G.~Montagna, O.~Nicrosini, G.~Passarino, F.~Piccinini and
R.~Pittau, Comput. Phys. Commun. 76 (1993) 328.}

\bibitem{yrwg}{D.~Bardin et al., {\it ``Electroweak Working Group
Report''}, in {\it ``Reports of the Working Group on Precision
Calculations for the Z Resonance''}, CERN Report 95-03 (Geneva, 1995), p.~7,
 edited by D.~Bardin, W.~Hollik and G.~Passarino.}

\bibitem{lettera}{G.~Montagna, O.~Nicrosini, G.~Passarino and F.~Piccinini,
Phys.~Lett. B335 (1994) 484.}

\bibitem{je}{F.~Jegerlehner, in Proceedings of the 1990 TASI in Elementary
Particle Physics, ed. by P.~Langacker and M.~Cvetic (World Scientific,
Singapore, 1991) p.476.}

%\bibitem{errori}{H.~J.~Behrend et al., Phys.~Lett. B183 (1987) 400;\\
%W.~de~Boer, Nucl.~Inst.~Meth. A278 (1989) 687.}

\bibitem{barb} {R.~Barbieri, M.~Beccaria, P.~Ciafaloni, G.~Curci
and A.~Vicer\'e, Phys. Lett. B288 (1992) 95; Nucl. Phys. B409 (1993) 105.}
\vskip 10pt

\bibitem{kat} A.~L.~Kataev, Phys. Lett. B287 (1992) 209.

\bibitem{aas} B.~A.~Kniehl, Nucl. Phys. B347 (1990) 86;
A.~Djouadi, Nuovo Cim. 100A (1988) 357.

\bibitem{qcd} K.~G.~Chetyrkin, Phys. Lett. B307 (1993) 169;
K.~G.~Chetyrkin and A.~Kwiatkowski, Phys. Lett. B305 (1993) 285;
S.~A.~Larin, T.~van~Ritbergen and J.~A.~M.~Vermaseren,
Phys. Lett. B320 (1994) 159;
K.~G.~Chetyrkin and J.~H.~K\"uhn, Phys. Lett. B308 (1993) 127.

\bibitem{ftjr} {J.~Fleischer, O.~V.~Tarasov, F.~Jegerlehner and P.~Raczka,
Phys.~Lett. B293 (1992) 437. See also: \\
G.~Buchalla and A.~J.~Buras, Nucl.~Phys.~B398 (1993) 285; \\
G.~Degrassi, Nucl.~Phys.~B407 (1993) 271.}

\bibitem{tar} L.~Avdeev, J.~Fleischer, S.~Mikhailov and O.~Tarasov, Phys.
Lett. B336 (1994) 560; \\
% Bielefeld preprint BI-TP-93-60; \\
A.~Sirlin, New York preprint NYU-TH-94/08/01 (August 1994).

\bibitem{hollik}{W.~F.~L.~Hollik, Fortschr. Phys. 38 (1990) 165. }

\bibitem{bkj}{F.~A.~Berends, R.~Kleiss and
S.~Jadach, Nucl. Phys. B202 (1982) 63.}

\bibitem{kkks} {B.~A.~Kniehl, M.~Krawczyk, J.~H.~K\"uhn and
R.~G.~Stuart, Phys.~Lett. B209 (1988) 337. See also: \\
S.~Jadach, M.~Skrzypek and M.~Martinez, Phys.~Lett. B280 (1992) 129.}
\vskip 10pt

\end{thebibliography}
\end{document}